\documentclass{aa}  

\usepackage{graphicx}

\usepackage{txfonts}

\usepackage[colorlinks, citecolor=blue]{hyperref}

\usepackage{natbib,twoopt}
\usepackage{tikz}

\graphicspath{{figures/}}

\newcommand{\mat}[1]{\vec{#1}}

\newcommand{\dmat}[1]{\mat{\dot{#1}}}
\newcommand{\body}{\mathcal{B}}
\newcommand{\Io}{{{\rm I}_\circ}}
\newcommand{\lagran}{{\cal L}}

\newcommand{\braket}[2]{\left\langle {#1}\,,\;{#2}\right\rangle}

\newcommand{\Id}{\mathbb{I}}
\newcommand{\norm}[1]{{\left\|{#1}\right\|}}
\newcommand{\dd}{\mathrm{d}}
\newcommand{\DD}[2]{\frac{\dd #1}{\dd #2}}
\newcommand{\deriv}[2]{\frac{\partial #1}{\partial #2}}
\newcommand{\jhat}{\mat{\hat{\mathcal{J}}}}
\newcommand{\comm}[1]{\left[ #1 \right]}

\newcommand{\moy}[1]{\overline{#1}}

\newcommand{\stat}{\mathrm{stat}}
\newcommand{\peri}{\mathrm{peri}}
\newcommand{\Frame}{\mathcal{F}}
\newcommand{\inertial}{\mathrm{inv}}
\newcommand{\orb}{\mathrm{orb}}
\newcommand{\Forb}{\Frame_\orb}
\newcommand{\Fin}{\Frame_\inertial}
\newcommand{\Fbody}{\Frame_\body}
\newcommand{\iorb}{{\vec i}}
\newcommand{\jorb}{{\vec j}}
\newcommand{\korb}{{\vec k}}
\newcommand{\ii}{\mathrm{i}}
\newcommand{\Rayleigh}{\mathcal{D}}
\newcommand{\osci}{\mathrm{peri}}
\newcommand{\core}{\mathrm{c}}
\newcommand{\mantle}{\mathrm{m}}
\newcommand{\Ioc}{{\rm I}_{\circ\core}}
\newcommand{\Iom}{{\rm I}_{\circ\mantle}}
\newcommand{\TRIP}{{\tt TRIP}}
\newcommand{\emi}{\varepsilon_{\mantle 1}}
\newcommand{\emj}{\varepsilon_{\mantle 2}}
\newcommand{\emk}{\varepsilon_{\mantle 3}}
\newcommand{\eci}{\varepsilon_{\core 1}}
\newcommand{\ecj}{\varepsilon_{\core 2}}

\newcommand{\IfIc}{\iota_\core}

\DeclareMathOperator{\trace}{Tr}
\DeclareMathOperator{\sign}{sign}

\begin{document}

   \title{Obliquity of Mercury with a fluid core and a deformable mantle}

   \subtitle{}

   \author{Valentin Etienne\inst{1}
          \and
          Gwena{\"e}l Bou{\'e}\inst{1}
          \and
          Clodoaldo Ragazzo\inst{2}
          }

   \institute{
              LTE, Observatoire de Paris, Université PSL, Sorbonne Université, Université de Lille, LNE, CNRS, 61 Avenue de l’Observatoire, 75014 Paris, France\\
              \email{valentin.etienne@obspm.fr}\\
              \email{gwenael.boue@obspm.fr}
              \and
              Instituto de Matem{\'a}tica e Estat{\'i}stica, Universidade de S{\~a}o Paulo, R. do Mat{\~a}o, 1010, S{\~a}o Paulo, SP 05508-090, Brazil\\
              \email{ragazzo@usp.br}
             }

  \abstract
{

In anticipation of BepiColombo's orbital insertion in 2026, this study revisits the rotation of Mercury, modelling it with a fluid core and a deformable mantle. For this well-established physical model, we provide an analytical solution based on a novel Lagrangian formalism. Our approach enables a comprehensive 3D solution for the mantle's orientation, including the phase in longitude at the perihelion. The model incorporates dissipative torques from core-mantle boundary friction and tidal deformations, conservative pressure torques, and the effect of the planet's apsidal precession.

We validate the equilibrium positions obtained numerically in previous studies, but identify discrepancies in earlier analytical works. These inconsistencies overestimated the out-of-Cassini-plane offset induced by tidal deformation by a factor of 100. Additionally, we explore the parameter space to determine where the spin axis deviates most from the Cassini plane and identify a region exhibiting bifurcation in the spin axis equilibrium position. This region cannot be adequately analysed using a model linearised with respect to the obliquity angle.

}

   \keywords{Mercury --
                Rotational dynamics --
                Tides --
                Solid body --
                Interiors
               }

   \maketitle

\section{Introduction}
The existence of a molten core in Mercury has been inferred since the detection of its magnetic field by the Mariner 10 spacecraft \citep{Ness74}. Shortly thereafter, \citet{Peale76} discussed methods to independently estimate the polar moments of inertia, $C$ for the entire planet and $C_\mantle$ for the mantle alone, using the planet's obliquity and forced libration, knowing the gravitational coefficients $C_{20}$ and $C_{22}$. This estimation is feasible due to the core-mantle coupling, which is weak enough for the core to avoid following the mantle's rapid 88-day libration period, yet strong enough for the core to follow the mantle's slow 327,000-year nodal precession \citep{Peale76}\footnote{Here, we quote a refined value of the nodal precession from \citep{Stark_rotation}}.

Subsequent measurements have been conducted using Earth-based radar observations for libration \citep{Margot07} and data from the MESSENGER spacecraft for obliquity and gravity coefficients \citep{Mazarico14}. These observations nicely confirm the predictions of \citet{Peale76}, with current estimates placing $C/(MR2)$ at $0.333\pm0.005$ and $C_\mantle/C$ at $0.443\pm0.019$ (\citealp{Genova19}; see also \citealp{Goossens22}).

With this value of the mantle's moment of inertia, the period of Mercury's free oscillation in longitude is close to Jupiter's 12-year orbital period \citep{Margot07}. The amplitude and phase of the forced libration at this period should thus allow for the refinement of the mantle's moment of inertia \citep{Peale09}. However, observations over a sufficiently long timespan are still missing to properly fit these two quantities \citep{Margot07}.

Other physical properties that can be probed by rotation include tidal dissipation and core-mantle interactions, such as viscous coupling, topographic coupling, electromagnetic coupling, and pressure coupling \citep{Peale14}. The pressure torque, related to the flattening of the core, is a conservative torque that mainly modifies the equilibrium obliquity \citep{Peale14}.

All other dissipative effects influence the amplitude and phase of the forced libration at 12 years \citep{Peale09}, and they also displace the spin axis from the plane containing the normal to the invariant plane and the orbit pole, known as the Cassini plane \citep{Peale14}.

The dissipative effects on Mercury's spin axis orientation have been analysed numerically by \citet{Peale14}, and tides in particular, characterised by the ratio $k_2/Q$ of the Love number $k_2$ divided by the quality factor $Q$, have been studied analytically for a solid Mercury without a molten core by \citet{Baland17}. Notably, all core-mantle dissipative mechanisms have quite similar signatures \citep{Peale14} and a significantly stronger effect than tides \citep{Peale05}. The addition of a solid inner core has been studied both numerically and analytically by \citep{Peale16,Dumberry21,MacPherson22}.

Concurrently, observational studies of Earth-based measurements and MESSENGER data have constrained Mercury's spin axis orientation, suggesting that Mercury's spin indeed deviates from the Cassini plane by a few arcseconds \citep{Margot12,Mazarico14,Stark_rotation,Stark_messenger,Verma16,Genova19,Konopliv20,Bertone21}. However, current observational uncertainties still leave room for further refinement, particularly in constraining the core-mantle interactions.

The upcoming BepiColombo mission, a joint endeavour between the European Space Agency (ESA) and the Japan Aerospace Exploration Agency (JAXA), launched on October 19, 2018, represents a significant advancement in our ability to probe Mercury’s geophysical properties. Upon its scheduled orbital insertion in 2026, BepiColombo will deploy its suite of state-of-the-art instruments to achieve unprecedented accuracy in measuring geodetic parameters such as the gravitational coefficients $C_{20}$ and $C_{22}$, the amplitude of longitudinal librations, the orientation of the spin axis, and tidal deformation \citep{Genova21}. These improved measurements will provide a more precise assessment of Mercury's rotational motion, offering constraints on its interior structure \citep{Steinbrugge21}.

In this study, we revisit Mercury's rotational dynamics and derive analytical formulae for the orientation of its spin axis, extending the model presented in \citep{Baland17} to include a fluid core. These formulae are designed to avoid the need for numerical integrations as in \citep{Peale14}. Given that the different core-mantle dissipative processes discussed by \citet{Peale14} have similar signatures, here, we only consider the viscous coupling at the core-mantle boundary (CMB), which is proportional to the relative angular velocity between the core and the mantle \citep[e.g.][]{Peale07}.

While acknowledging the comprehensive model of \citet{MacPherson22} that includes the solid core, our initial focus is on refining the formula for the deviation of the spin axis from the Cassini plane induced by tides, addressing an inconsistency identified in \citep{Baland17} and subsequently reproduced in \citep{MacPherson22}. Additionally, we highlight a non-linear effect that emerges at low viscosity for specific values of the pressure torque, associated with the substantial tilt of the core's angular velocity as depicted in \citep[Fig.~3]{Peale14} and discussed in \citep{Boue20}. The exploration of more realistic interior structures is reserved for future work.

Our analysis employs the Lagrangian formalism developed in \citep{Ragazzo15} and \citep{Boue20}, which is mathematically equivalent to the angular momentum approach followed by \citet{Baland17}. In Section~\ref{sec.rigid} and \ref{sec.deformable}, we pursue a similar series of approximations as in \citep{Baland17} to obtain analytical solutions. Section \ref{sec.rigid} is dedicated to the rigid body approximation while the tidal deformation is accounted in Section~\ref{sec.deformable}. The key distinction, beyond the incorporation of the fluid core in Section~\ref{sec.core}, lies in our consideration of the complete 3D rotation matrix defining Mercury's orientation (three degrees of freedom), whereas \citet{Baland17} focused solely on the spin axis (two degrees of freedom). As demonstrated in our results in Section~\ref{sec.results}, this methodological difference significantly impacts the outcomes, despite both approaches yielding consistent results when considering only the projection of the torque orthogonal to the spin axis. Our conclusions are presented in Section~\ref{sec.conclusion}.

In this paper, we follow the notation of \citep{Boue20} and represent each vector $\hat v$ by a skew-symmetric matrix $\mat v$ such that
\begin{equation}
\hat v = \begin{pmatrix} x \\ y \\ z \end{pmatrix} \rightarrow \vec v = \begin{pmatrix} 0 & -z & y \\z & 0 & -x \\ -y & x & 0 \end{pmatrix}\,.
\end{equation}
The dot product $\braket{}{}$ of two matrices is defined as
\begin{equation}
\braket{\vec a}{\vec b} \equiv \frac{1}{2}\trace\left(\vec a^T\,\vec b\right) = \hat a \cdot \hat b\,,
\end{equation}
where $\trace$ denotes the trace operator, and the norm $\sqrt{\braket{\vec v}{\vec v}}$ of a vector $\vec v$ is denoted $\norm{\vec v}$.

\section{Mercury as a rigid body}
\label{sec.rigid}
Here, we retrieve the results of \citet{Baland17} using the formalism of \citep{Boue20}. The goal is to introduce the methodology on a simple case before adding complexity, namely deformation and a liquid core.

For the sake of generality, hereafter we refer to Mercury simply as a body $\body$ of mass $M$ and radius $R$ orbiting a central point mass object $m_0$.  Let $\mat R_\body$ be the rotation matrix between an inertial reference frame $\Frame_\inertial$ and a frame $\Fbody$ attached to $\body$. For the inertial frame, we take a basis $(\mat i_\inertial, \mat j_\inertial, \mat k_\inertial)$ whose third axis is normal to the invariant plane with respect to which the orbit is assumed to have a constant inclination $i$. As for the body fixed frame, we choose the basis composed of the principal axes of inertia $(\mat I, \mat J, \mat K)$, in which the moments of inertia are denoted $A$, $B$ and $C$ along the first, second and third axis, respectively.

Let $\Omega$ be the longitude of the ascending node of the orbit reckoned from the invariant plane. We assume the planet to be in a $p$\,:1 spin-orbit resonance, meaning that its angular velocity with respect to the inertial frame is
\begin{equation}
\mat\omega_\body = \dot\Omega\mat k_\inertial + \omega_p\vec K\,,\qquad
\omega_p \equiv pn + \dot\varpi - \dot\Omega\,,
\end{equation}
with $n=\sqrt{Gm_0/a^3}$ the orbital mean motion and $\varpi$ the longitude of the periapsis. In practice, for Mercury, we have $p =3/2$. In the expression of the mean motion, $G$ is the gravitational constant and $a$ the semi-major axis.

We consider that the position of the Cassini state corresponds to a fixed orientation of the vector $\mat K$ in a frame rotating at the orbital nodal precession frequency. Note that in the literature, Cassini states are sometimes defined using the angular momentum vector \citep{Noyelles11} instead of the body polar axis. However, both axes are very close, and for our purpose the two definitions are equivalent. Let us then introduce the frame $\Forb$ attached to the orbit whose basis $(\iorb, \jorb, \korb)$ is such that $\iorb$ is along the orbital ascending node with respect to the invariant plane and $\korb$ is along the orbital angular momentum. The transformation of coordinates between the orbital frame $\Forb$ and the inertial invariant frame $\Fin$ is given by $\mat R_\inertial = \mat R_3(\Omega)\mat R_1(i)$ where $\mat R_1$ denotes a rotation around the first axis and $\mat R_3$ around the third axis\footnote{The rotation matrices are defined as
$$
\mat R_1(\theta) = \begin{pmatrix}
1 & 0 & 0 \\ 0 & \cos\theta & -\sin\theta \\ 0 & \sin\theta & \cos\theta
\end{pmatrix}\,,\qquad
\mat R_3(\theta) = \begin{pmatrix}
\cos\theta & -\sin\theta & 0 \\ \sin\theta & \cos\theta & 0 \\ 0 & 0 & 1
\end{pmatrix}\,.
$$
}. The rotation matrix between the body fixed frame $\Fbody$ and the inertial frame $\Fin$ can itself be decomposed as follows
\begin{equation}
\label{eq.RBody}
\mat R_\body = \mat R_3(\Omega)\, \mat R_1(i)\, \mat R\, \mat R_3(\theta_p)\,,
\end{equation}
with $\theta_p = pM + \varpi - \Omega + \gamma_0$ the rotation angle of the planet, $\gamma_0$ an offset being equal to 0 or $\pi/2$ depending on the resonance \citep[][page 139]{Wisdom84}, $M$ the mean anomaly, and $\mat R$ the rotation defining the orientation of the Cassini states. Were the body not in a Cassini state, the expression of $\mat R_\body$ would still hold, but the angular velocity would have an extra term $\vec \omega = \dmat R\mat R^T$ accounting for the time derivative of $\mat R$, namely
\begin{equation}
\vec\omega_\body = \dot\Omega\mat k_\inertial + \vec \omega + \omega_p \vec K\,.
\end{equation}

In the following, we provide the equations of motion in terms of the variables $(\vec\omega, \mat R)$, the Cassini state being an equilibrium of the secular problem (averaged over the orbital mean anomaly) corresponding to $\vec\omega = \vec 0$ and $\mat R = \mathrm{constant}$. When deformation is taken into account, the matrix of inertia (or equivalently the traceless symmetric $\mat B$-matrix, as defined in \citep{Ragazzo15}) is allowed to vary, and an equation of motion is added for this variable.

The Lagrangian describing the motion of $(\vec\omega, \mat R)$, written in the orbital frame $\Forb$ is \citep[e.g.][]{Boue20}
\begin{equation}
\lagran =  \frac{\Io}{2}\left(\norm{\vec\omega_\body}^2
 - 2 \braket{\mat R \mat B' \mat R^T}{\vec\omega_\body^2}\right)
+ 3 \Io\frac{Gm_0}{a^3} \braket{\mat R \mat B' \mat R^T}{\mat S}\,,
\end{equation}
with $\Io\equiv (A+B+C)/3$ the mean moment of inertia of the body $\body$, $\mat B' = \mat R_3(\theta_p)\mat B\mat R_3(-\theta_p)$, the $\mat B$-matrix written in the precessional frame as defined in \citep[][Sect.~7]{Ragazzo22} such that $\mat R \mat B' \mat R^T$ is expressed in the orbital frame $\Forb$, and $\mat S$ a traceless symmetric matrix encapsulating the orbital dependency of the spin-orbit potential energy. We only consider Mercury's gravity field at second order, as the higher orders measured by MESSENGER \citep{Smith12,Mazarico14} have negligible effects on the orientation of the spin axis \citep{Noyelles13,Baland17}. We decompose $\mat B$ and $\mat S$ onto real spherical harmonic matrices as defined in \citep{Boue20}, namely,
\begin{align}
\mat B =&\, \frac{2}{3}\frac{MR^2}{\Io} \left(C_{20}\mat Y_{20} + C_{22} \mat Y_{22}\right)\,,\qquad \\
\mat S =&\, \frac{a^3}{r^3}\bigg(
-\frac{1}{3}\mat Y_{20} + \frac{1}{6}\Big(\cos(2\varv+2\varpi-2\Omega)\mat Y_{22} \nonumber \\
& + \sin(2\varv+2\varpi-2\Omega)\mat Y_{2-2}\Big)\bigg)\,,
\end{align}
where $r$ is the orbital separation and $\varv$ the true anomaly. Parameters $C_{20}$ and $C_{22}$ are Stokes coefficients related to the moments of inertia according to
\begin{align}
C_{20} \equiv&\; \frac{A+B-2C}{2MR^2}\,, \\
C_{22} \equiv&\; \frac{B-A}{4MR^2}\,.
\end{align}
The equations of motion are \citep{Boue20}
\begin{equation}
\label{eq.EulerLagrangeRigid}
\DD{}{t}\deriv{\lagran}{\mat\omega} = \comm{\vec\omega, \deriv{\lagran}{\vec\omega}} + \jhat(\lagran)\,,\qquad
\dmat R = \vec\omega\mat R\,,
\end{equation}
where $\jhat$ is the spin operator defined for all function $f(\mat R)$ as $\jhat(f) \equiv (\partial f / \partial \mat R)\mat R^T$ and $[\,,\,]$ is the usual commutator of two matrices : $\comm{\mat A, \mat B} \equiv \mat A\mat B - \mat B\mat A$. The equilibrium is reached when the body's angular momentum
\begin{equation}
\vec\pi \equiv \deriv{\lagran}{\vec \omega} = \Io \left(\vec\omega_\body + \mat R \mat B'\mat R^T \vec \omega_\body +
                 \vec\omega_\body\,\mat R\mat B'\mat R^T\right)
\label{eq.pi}
\end{equation}
averaged over an orbital period remains constant for $\vec\omega = \vec 0$. Substituting the expression of the Lagrangian in the equation of motion, one gets
\begin{equation}
\dmat\pi = -\dot\Omega\comm{\vec k_\inertial, \vec \pi} - 3 \Io\,n^2 \comm{\mat R \mat B' \mat R^T,\,\mat S}\,.
\label{eq.dotpisolid}
\end{equation}
Setting the time average of the body's angular momentum rate $\dmat \pi$ equal to $\vec 0$, we obtain an equation whose unknown is the rotation matrix $\mat R$ defining the orientation of the Cassini state in the orbital frame $\Forb$. In the case of Mercury, the deviation of the figure-axis $\vec K$ from the orbit normal $\vec k$ is very small (about 2 arcmin). It is thus legitimate to linearise Eq.~(\ref{eq.dotpisolid}) over the tilt angle $\vec \varepsilon = \varepsilon_1\vec i + \varepsilon_2\vec j + \varepsilon_3\vec k$, where $\varepsilon_1$ represents the in-plane offset of the Cassini state, $\varepsilon_2$ the deviation from the Cassini plane, and  $\varepsilon_3$ the rotation lag of the body around its spin axis or, equivalently, the centre of libration in longitude. With this approximation, $\mat R \approx \Id + \vec\varepsilon$ where $\Id$ is the identity matrix and $\mat R\mat B'\mat R^T \approx \mat B' + \comm{\mat \varepsilon, \mat B'}$. Moreover, the precession rate of the orbit $\dot\Omega$ is much smaller than the orbital mean motion $n$ in absolute value ($|\dot\Omega/n| \approx 7\times10^{-7}$), therefore we can safely ignore terms in $\dot\Omega^2$ before those in $n^2$. Applying the two simplifications in the expression of $\dmat\pi$ (Eq.~\ref{eq.dotpisolid}) and averaging over the mean anomaly, one gets
\begin{equation}
\label{eq.systsolid}
\begin{aligned}
&
\begin{bmatrix}
D & 0 & 0 \\
0 & D & 0 \\
0 & N & -4\kappa_{22}
\end{bmatrix} \begin{pmatrix} \varepsilon_1 \\ \varepsilon_2 \\ \varepsilon_3 \end{pmatrix} 
= -
\begin{pmatrix} 
N \\ 0 \\ 0
\end{pmatrix}
\\ &\qquad
+
\kappa_\omega \begin{pmatrix}
\cos(2\varpi-2\Omega) & \sin(2\varpi-2\Omega) & 0 \\
\sin(2\varpi-2\Omega) &-\cos(2\varpi-2\Omega) & 0 \\
0 & 0 & 0
\end{pmatrix}\begin{pmatrix} \varepsilon_1 \\ \varepsilon_2 \\ \varepsilon_3 \end{pmatrix} 
\end{aligned}
\end{equation}
with
\begin{equation}
D = \kappa + C\omega_p\,\dot\Omega\cos i\,,
\label{eq.D}
\end{equation}
and
\begin{equation}
N = \dot\Omega\sin i \left(C\omega_p-MR^2C_{20}\,\dot\Omega\cos i\right)\,.
\label{eq.N}
\end{equation}
The polar moment of inertia is given by $C = \Io - \frac{2}{3}\,MR^2C_{20}$. In the above expressions, we applied a notation similar to that introduced in \citep{Baland17} up to a factor $\frac{3}{2}n$. More explicitly, we have 
\begin{align}
\label{eq.k20}
\kappa_{20} =&\; -\frac{3}{2} n^2 MR^2 C_{20}\,X^{-3,0}_0(e) \,, \\
\label{eq.k22}
\kappa_{22} =&\; 3\,\sigma_p\,n^2 MR^2 C_{22}\,X^{-3,2}_{2p}(e)\,,
\end{align}
and
\begin{equation}
\kappa_\omega = 3\,\sigma_p\,n^2 MR^2 C_{22}\,X^{-3,0}_{2p}(e)\,,
\label{eq.kom}
\end{equation}
where the $X^{n,m}_k(e)$ are Hansen coefficients of the eccentricity $e$ defined such that
\begin{equation}
\left(\frac{r}{a}\right)^n \exp\left(\ii\, m \varv\right) \equiv \sum_{k=-\infty}^\infty X^{n,m}_k(e)\,\exp\left(\ii\, k M\right)\,,
\end{equation}
and $\sigma_p \equiv \cos(2\gamma_0) = \sign(X^{-3,2}_{2p}(e))$ \citep{Wisdom84}.
The system of equations (\ref{eq.systsolid}) was obtained with the help of the computer algebra system \TRIP{} \citep{Gastineau11}. This system includes trigonometric terms in $2\varpi-2\Omega$ due to the apsidal precession of the orbit as in \citep[][Eqs. (34, 35)]{Baland17}. These terms are responsible for a nutation of the spin axis at a slow period of about 84 kyr \citep{Baland17}.
To solve Eq.~(\ref{eq.systsolid}), we proceeded as in \citep{Baland17}, i.e. we first neglected the contribution due to the apsidal precession of the orbit, which amounts to set $\kappa_\omega \sim e^{2p}$ equal to zero. 
In that case, we get a first order solution $\varepsilon_1^{(1)} = \varepsilon_\Omega$ and $\varepsilon_2^{(1)} = \varepsilon_3^{(1)} = 0$ with
\begin{align}
\varepsilon_\Omega = -\frac{N}{D} = -\frac{\dot\Omega\sin i\,(C\omega_p-MR^2C_{20}\,\dot\Omega\cos i)}{\kappa + C\omega_p\,\dot\Omega\cos i}\,.
\label{eq.eps_Om_solid}
\end{align}
In a second step, we inserted this first order solution in the right-hand side of Eq.~(\ref{eq.systsolid}) and solved again the system. The result is 
\begin{subequations}%
\label{eq.solsolid}
\begin{align}
\varepsilon_1 &= \varepsilon_\omega\cos(2\varpi-2\Omega) + \varepsilon_\Omega\,, \\
\varepsilon_2 &= \varepsilon_\omega\sin(2\varpi-2\Omega)\,, \\
\varepsilon_3 &= \frac{N}{4\kappa_{22}}\varepsilon_\omega\sin(2\varpi-2\Omega)\,,
\end{align}%
\end{subequations}
with
\begin{equation}
\varepsilon_\omega = \frac{\varepsilon_\Omega \kappa_\omega }{\kappa + C\omega_p\,\dot\Omega\cos i}\,.
\label{eq.eps_om_solid}
\end{equation}

The results of this calculation from Eq.~(\ref{eq.eps_Om_solid}) to Eq.~(\ref{eq.eps_om_solid}) are very similar to those in \citep{Baland17}. The main difference is the libration angle $\varepsilon_3$ that is absent in \citep{Baland17}.

Let us rewrite the relevant formulae neglecting the nodal precession frequency $|\dot\Omega|$ before the mean motion $n$ in a simpler form~:
\begin{align}
\epsilon_\Omega &\approx \frac{C}{MR^2}\,\frac{p|\dot\Omega|}{n}\,\frac{\sin i}{-\frac{3}{2}C_{20}X_0^{-3,0}(e)+3\sigma_pC_{22}X_{2p}^{-3,2}(e)}\,, \\
\epsilon_\omega &\approx \frac{2\sigma_pC_{22}X_{2p}^{-3,0}(e)}{-C_{20}X_0^{-3,0}(e) + 2\sigma_pC_{22}X_{2p}^{-3,2}(e)}\varepsilon_\Omega\,.
\end{align}

\section{Mercury as a deformable body}
\label{sec.deformable}
Here, we consider the case where the body is deformable. As in \citep{Ragazzo15}, we introduce an effective macroscopic rigidity $\mu$ accounting for the elastic response of the body as a whole according to the elastic potential energy
\begin{equation}
U_\mathrm{el} = \frac{\Io}{2}\mu\,\norm{\mat B-\mat B_0}^2\,,
\end{equation}
where $\mat B_0$ is a permanent deformation induced by a prestress \citep{Ragazzo22}. The effective rigidity $\mu$ is related to the Love number $k_2$ and to the tidal parameter \citep[Eq.~(B.12)]{Baland17}
\begin{equation}
q_t = -3\frac{Gm_0}{GM}\left(\frac{R}{a}\right)^3
\end{equation}
through
\begin{equation}
\frac{1}{\mu} = -\frac{k_2q_t}{9n^2}\frac{MR^2}{\Io}\,.
\end{equation}
For a smooth comparison with \citep{Baland17}, we also introduce the ratio of the centrifugal acceleration to the gravitational acceleration $q_r = -\frac{1}{3}p^2q_t$\,.
With the elastic potential energy, the Lagrangian of the problem now reads
\begin{equation}
\label{eq.LagrangeDeformable}
\begin{aligned}
\lagran =&  \frac{\Io}{2}\left(\norm{\vec\omega_\body}^2
 + 2 \braket{\mat R \mat B' \mat R^T}{\vec\omega_\body \vec\omega_\body^T}\right)
 - \frac{\Io}{2}\mu\,\norm{\mat B - \mat B_0}^2 \\ &
+ 3 \Io\frac{Gm_0}{a^3} \braket{\mat R \mat B' \mat R^T}{\mat S}\,.
\end{aligned}
\end{equation}
The viscous deformation of the body as a whole is controlled by a Rayleigh dissipation function $\Rayleigh$ that is a function of an effective macroscopic viscosity $\eta$ such that
\begin{equation}
\label{eq.Rayleigh}
\Rayleigh = \frac{\Io}{2} \eta\,\norm{\dmat B}^2\,.
\end{equation}
From the effective macroscopic viscosity $\eta$ and the effective macroscopic rigidity $\mu$, one can define a Maxwell relaxation time $\tau_M \equiv \eta/\mu$ that is related to the tidal quality factor $Q = 1/(n\tau_M)$. Introducing the phase shift $\zeta \equiv 1/Q = n\tau_M$ as in \citep{Baland17}, we get the relation
\begin{equation}
\eta = -\frac{9n\zeta}{k_2q_t}\frac{\Io}{MR^2}\,.
\end{equation} 
The equations of motion (\ref{eq.EulerLagrangeRigid}) for $(\vec\omega, \mat R)$ still hold. We now add the equation describing the evolution of the $\mat B$-matrix, namely,
\begin{equation}
\DD{}{t}\deriv{\lagran}{\dmat B} = \deriv{\lagran}{\mat B} - \deriv{\Rayleigh}{\dmat B}\,.
\label{eq.Bmotion}
\end{equation}
Since we discard the kinetic energy associated with the deformation of the body (there is no $\dmat B$ dependency in the Lagrangian), the equation of motion (\ref{eq.Bmotion}) is equal to $\mat 0$. This approximation is justified in \citep{Correia18}. Let $\delta\mat B \equiv\mat B-\mat B_0$ be the deviation of the $\mat B$-matrix from its relaxed state $\mat B_0$. From the expression of the Lagrangian (\ref{eq.LagrangeDeformable}) and of Rayleigh dissipation function (\ref{eq.Rayleigh}), the equation of motion (\ref{eq.Bmotion}) reads
\begin{equation}
\delta\mat B + \frac{\eta}{\mu}\delta\dmat B = \frac{1}{\mu}\mat F
\label{eq.explicitBmotion}
\end{equation}
with $\mat F = \mat F_r + \mat F_t$ and
\begin{align}
\label{eq.Fr}
\mat F_r =&\; \mat R_3(-\theta_p)\mat R^T\left(\vec\omega_\body\vec\omega_\body^T\right)\mat R\mat R_3(\theta_p) - \frac{2}{3}\norm{\vec\omega_\body}^2\Id\,, \\
\mat F_t =&\; 3n^2 \mat R_3(-\theta_p)\mat R^T\mat S\mat R\mat R_3(\theta_p)\,.
\label{eq.Ft}
\end{align}
Given that the influence of deformation on the spin axis orientation is weak, only the leading-order terms are retained in the evaluation of the `force' $\mat F$ on the left-hand side of (\ref{eq.explicitBmotion}). Consequently, we disregard the obliquity, assuming the spin axis to be perpendicular to the orbital plane, and neglect the precession frequency relative to the spin rate. With this approximation,
\begin{align}
\mat F_r \approx&\; \omega_p^2 \left(\vec k\vec k^T - \frac{2}{3}\Id\right)\,, \\
\mat F_t \approx&\; 3n^2 \mat R_3(-\theta_p)\mat S\mat R_3(\theta_p)\,.
\end{align}
The static part $\mat F^\mathrm{stat} \equiv \moy{\mat F}$ of $\mat F$ generates a constant deformation $\delta\mat B^\mathrm{stat}$ for which the Love number $k_2$ admits the fluid limit $k_f$. We get
\begin{equation}
\delta\mat B^\mathrm{stat} = \frac{2}{3}\bigg(\delta C_{20}^\mathrm{stat}\,\mat Y_{20} + \delta C_{22}^\mathrm{stat}\,\mat Y_{22}\bigg)
\end{equation}
with
\begin{align}
\delta C_{20}^\mathrm{stat} =&\;\frac{k_f}{6}\left(-2q_r + q_t X^{-3,0}_0(e)\right)\,, \\
\delta C_{22}^\mathrm{stat} =&\;-\frac{k_f}{12}q_tX^{-3,2}_{2p}(e) \cos(2\gamma_0)\,.
\end{align}
These two formulae are in perfect agreement with \citep[Eqs.~(B.13-B.14)]{Baland17}. Putting numbers into these expressions show that the hydrostatic parts $\delta C_{20}^\mathrm{stat}$ and $\delta C_{22}^\mathrm{stat}$ represent about 1\% of the observed gravity field coefficients $C_{20}$ and $C_{22}$, respectively.

We now turn to the oscillatory part $\mat F - \moy{\mat F}$ that produces a significant out of plane offset of the Cassini state of about 1 arcsec according to \citep{Baland17}. At first order in $\tau_M = \eta/\mu$, the oscillatory (or periodic) part $\delta\mat B^\osci$ of $\delta \mat B$ is
\begin{equation}
\delta\mat B^\osci = \frac{1}{\mu}\left(\mat F - \moy{\mat F} - \tau_M \dmat F\right)\,,
\label{eq.Bosci}
\end{equation}
with $\mu$ evaluated at $k_2 \neq k_f$. In our approximation, the centrifugal force remains static in the body frame. Consequently, we can compute the periodic component of $\delta \mat B$ using Eq.~(\ref{eq.Bosci}), substituting the force $\mat F$ with $\mat F_t$.

Once $\delta\mat B^\osci$ is known, the system (\ref{eq.systsolid}) is recomputed with $\mat B'$ replaced by $\mat B' + \mat R_3(\theta_p)\delta\mat B^\osci\mat R_3(-\theta_p)$ in Eqs.~(\ref{eq.pi} and \ref{eq.dotpisolid}). The result is averaged over the mean anomaly and over the longitude of the periapsis to focus on the out of plane offset of the Cassini state. The calculation done with \TRIP{} gives
\begin{equation}
\label{eq.systdeform}
\begin{bmatrix}
D & 0 & 0 \\
0 & D & 0 \\
0 & N & -4\kappa_{22}
\end{bmatrix} \begin{pmatrix} \delta\varepsilon_1 \\ \delta\varepsilon_2 \\ \delta\varepsilon_3 \end{pmatrix} 
= -
\begin{pmatrix}
\kappa_{\zeta,c}\,\varepsilon_\Omega \\
\kappa_{\zeta,s}\,\varepsilon_\Omega \\
\kappa_{\zeta,s}+\kappa_{\zeta,n}
\end{pmatrix}\,,
\end{equation}
where $\delta\vec\varepsilon = (\delta\varepsilon_1, \delta\varepsilon_2, \delta\varepsilon_3)$ is a correction to the solution $\vec\varepsilon = (\varepsilon_1, \varepsilon_2, \varepsilon_3)$ obtained in the rigid case (Eq.~\ref{eq.solsolid}). The constants in this new system are defined as
\begin{align}
\kappa_{\zeta,c} \equiv&\; \frac{1}{4}k_2q_tn^2\,MR^2\left(\left(X^{-3,2}_{2p}(e)\right)^2 + \left(X^{-3,0}_0(e)\right)^2\right)\,,\\
\kappa_{\zeta,s} \equiv&\; \frac{1}{2}k_2q_t\zeta n\omega_p\,MR^2\,X^{-6,0}_0(e)\,, \\
\kappa_{\zeta,n} \equiv&\; \frac{1}{2}k_2q_t\zeta n\,MR^2\left(\omega_pX^{-6,0}_0(e)-2n\sqrt{1-e^2}X^{-8,0}_0(e)\right)\,.
\end{align}
The first line of (\ref{eq.systdeform}) provides a small correction to the in-plane offset of the Cassini state $\varepsilon_\Omega$ of the order of $0.1\%$ that becomes
\begin{equation}
 \varepsilon_\Omega^{k_2} = \varepsilon_\Omega\left(1-\frac{\kappa_{\zeta,c}}{D}\right)\,.
\label{eq.eps_Om_k2}
\end{equation}
The third line corresponds to a rotation of the body around its figure axis of about 3 arcsec, and the second line gives the out of plane offset of the Cassini state, namely, $\delta\varepsilon_2 = \varepsilon_\zeta$ with
\begin{equation}
\varepsilon_\zeta = -\frac{\kappa_{\zeta,s}\varepsilon_\Omega}{D} 
                  = -\frac{\kappa_{\zeta,s}\varepsilon_\Omega}{\kappa + C\omega_p\dot\Omega\cos i}\,.
\label{eq.eps_zeta}
\end{equation}
This result shows some similarities with \citep[Eq.~(66)]{Baland17}. Our denominator is directly expressed in terms of the entries of the $\mat B$-matrix while in \citep{Baland17} the same denominator (up to a coefficient $\cos i$) explicitly shows the contribution of the static hydrostatic adjustment (related to the $\delta\mat B^\mathrm{stat}$-matrix) with respect to a situation where only the relaxed gravity field coefficients (related to the $\mat B_0$-matrix) would have been taken into account. But since observations cannot distinguish these two constant components, we chose not to split the entries of the $\mat B$-matrix (which means that in our notation, $C$, $C_{20}$ and $C_{22}$ do include the permanent hydrostatic deformation).

The primary difference between our result (\ref{eq.eps_zeta}) and \citep[Eq.~(66)]{Baland17} lies in the numerator. Our expression includes only the first term of \citeauthor{Baland17}'s numerator, omitting the $\cos i$ factor. To derive their Eq.~(66), \citet{Baland17} solved two components of their 3D angular momentum equation (Eq.~(59)), where the unknown is the spin axis ($\hat s(t)$ in their notation). The challenge arises because their Eq.~(59) for $\dd \hat s/\dd t$ is not orthogonal to $\hat s$, despite $\hat s$ being a unit vector. By definition, if $\hat s\cdot\hat s = 1$, then $\DD{\hat s}{t}\cdot\hat s$ should be zero.

To address this issue, \citet{Baland17} would have considered the two components of the torque orthogonal to $\hat s$ (or to the orbit pole $\vec k$, given the very small obliquity of the Cassini state $\varepsilon_\Omega$). They instead chose the two components in the invariant plane, resulting in their numerator containing the second component of the right-hand side of our Eq.~(\ref{eq.systdeform}) multiplied by $\cos i$, plus the third component of the right-hand side of our Eq.~(\ref{eq.systdeform}) multiplied by $\sin i$. As a result, the reasoning used in \citep{Baland17} to obtain their final expression Eq.~(66) is not correct. We nevertheless verified that our expression for $\varepsilon_\zeta$ can be derived from \citeauthor{Baland17}'s equation of motion (Eq.~(59)) using the correct projections.

\section{Influence of a fluid core}
\label{sec.core}
In this section, we take into account the presence of a molten core in the interior of Mercury. The mantle surrounding the core is still subject to tidal deformation. The fluid core is parameterised by its mean moment of inertia $\Ioc$ and by the $\mat B$-matrix \begin{equation}
\mat B_\core \equiv \frac{2}{3}\frac{MR^2}{\Ioc} \left(C_{20}^{\,\core}\,\mat Y_{20} + C_{22}^{\,\core}\,\mat Y_{22}\right)\,,
\label{eq.Bcore}
\end{equation}
where the Stokes coefficients $C^{\,\core}_{20}$ and $C^{\,\core}_{22}$ are normalised by the product $MR^2$ of the total mass $M$ and the full radius squared $R^2$ of the body. The matrix $\mat B_\core$ implicitly defines the shape of the core mantle boundary (CMB). The mantle is equivalently characterised by its mean moment of inertia $\Iom$ and the $\mat B$-matrix 
\begin{equation}
\mat B_\mantle \equiv \frac{2}{3}\frac{MR^2}{\Iom} 
                      \left(C_{20}^{\,\mantle}\,\mat Y_{20} + C_{22}^{\,\mantle}\,\mat Y_{22}\right)\,,
\end{equation}
where $C^{\,\mantle}_{20}$ and $C^{\,\mantle}_{22}$ also are normalised by $MR^2$.
The overall body has a mean moment of inertia $\Io = \Ioc + \Iom$ and its gravity field is represented by the matrix $\mat B$ such that $\Io\mat B = \Ioc\mat B_\core + \Iom\mat B_\mantle$. These three matrices rotate in concert with the rotation matrix of the mantle $\mat R$. Under tidal perturbation, $\mat B_\core$ and $\mat B_\mantle$ could vary independently. We should expect two normal modes of vibration, a symmetric one where the two matrices oscillate in phase and another antisymmetric where they oscillate in antiphase. For the sake of simplicity, we only consider the symmetric mode and assume that each of these $\mat B$-matrices remain proportional to the total $\mat B$-matrix with a constant coefficient that we denote $f_\core$ for the core and $f_\mantle$ for the mantle, namely,
\begin{equation}
\mat B_\core \equiv f_\core \mat B\,,\qquad\mathrm{and}\qquad
\mat B_\mantle \equiv f_\mantle \mat B\,.
\end{equation}
By definition of the matrix $\mat B$, the two coefficients $f_\core$ and $f_\mantle$ are related to each other by the formula
\begin{equation}
\Ioc f_\core + \Iom f_\mantle = \Io\,.
\end{equation}
The rotational state of the core is given by its angular velocity 
\begin{equation}
\vec\omega_\body^\core = \dot\Omega \vec k_\inertial + \vec\omega_\core
\end{equation}
where $\vec\omega_\core$ is the rotation speed with respect to the orbital frame. For the mantle, we chose the same decomposition as in the single layer model above, i.e. we define its angular velocity as
\begin{equation}
\vec\omega_\body^\mantle = \dot\Omega \vec k_\inertial + \vec\omega_\mantle + \omega_p\vec K\,.
\end{equation}
The variables of the problem are $(\mat \omega_\core, \mat \omega_\mantle, \mat R, \mat B)$.
At a Cassini state, the system is in a relative equilibrium characterised by $\mat \omega_\mantle = \mat 0$, $\mat\omega_\core = $ constant and $\mat R = $ constant. 

The Lagrangian of the system is the kinetic energy of the core plus the kinetic energy of the mantle minus the spin-orbit potential energy minus the elastic potential energy of deformation
\begin{equation}
\begin{aligned}
\lagran &= \frac{\Ioc}{2}\left(\norm{\vec\omega_\body^\core}^2 + 2f_\core\braket{\mat R\mat B'\mat R^T}{\vec\omega_\body^\core\vec\omega_\body^{\core T}}\right)
\\ & + \frac{\Iom}{2}\left(\norm{\vec\omega_\body^\mantle}^2 + 2f_\mantle\braket{\mat R\mat B'\mat R^T}{\vec\omega_\body^\mantle\vec\omega_\body^{\mantle T}}\right)
\\ & + 3\Io n^2\braket{\mat R\mat B'\mat R^T}{\mat S}
 - \frac{\Io}{2}\mu\,\norm{\mat B-\mat B_0}^2\,.
\end{aligned}
\end{equation}
To this Lagrangian, we added a Rayleigh dissipation function similar to the one used above in the single layer approach but with an extra term accounting for a viscous friction at the CMB, namely,
\begin{equation}
\Rayleigh = \frac{\Io}{2}\eta\,\norm{\dmat B}^2 + \frac{\Io}{2}\nu\,\norm{\vec\omega_\body^\core - \vec\omega_\body^\mantle}^2\,,
\end{equation}
where $\nu$ (not to be confused with the true anomaly $\varv$) has the dimension of a frequency related to the inverse of the characteristic timescale
\begin{equation*}
\tau_\nu = \left(1-\frac{C_\mantle}{C}\right)\frac{C_\mantle}{C}\frac{1}{\nu}
\end{equation*}
required to damp the relative angular velocity of the core with respect to that of the mantle \citep{Peale09}.
In this formula, $C_\mantle = \Iom - \frac{2}{3} f_\mantle MR^2 C_{20}$ is the mantle's polar moment of inertia.
The equations of motion are
\begin{subequations}%
\begin{align}%
\dmat\pi_\core =&\; \comm{\vec\omega_\core, \vec\pi_\core} - \deriv{\Rayleigh}{\vec\omega_\core}\,, \\
\dmat\pi_\mantle =&\; \comm{\vec\omega_\mantle, \vec\pi_\mantle} - \deriv{\Rayleigh}{\vec\omega_\mantle} - \jhat(\lagran)\,, \\
\vec 0 =&\; \deriv{\lagran}{\mat B} - \deriv{\Rayleigh}{\dmat B}\,,
\label{eq.formalB}
\end{align}%
\label{eq.eq_motion_2layers}%
\end{subequations}
where the angular momenta are given by
\begin{align}
\vec\pi_\core \equiv&\; \deriv{\lagran}{\vec\omega_\core} 
 = \Ioc\left(\vec\omega_\body^\core 
   + f_\core\left(\mat R \mat B'\mat R^T\vec\omega_\body^\core 
   + \vec\omega_\body^\core \mat R\mat B'\mat R^T\right)\right)\,, \\
\vec\pi_\mantle \equiv&\; \deriv{\lagran}{\vec\omega_\mantle} 
 = \Iom\left(\vec\omega_\body^\mantle 
   + f_\mantle\left(\mat R \mat B'\mat R^T\vec\omega_\body^\mantle 
   + \vec\omega_\body^\mantle \mat R\mat B'\mat R^T\right)\right)\,.
\end{align}
To get the position of the Cassini state, we proceed as above. The periodic oscillations $\delta \mat B^\peri$ of the $\mat B$-matrix are computed at the leading order and injected in the first two equations of motion whose explicit expressions read
\begin{align}
\label{eq.dotpic}
\dmat\pi_\core =&\; -g\comm{\vec k_\inertial, \vec\pi_\core} - \Ioc f_\core\comm{\mat R\mat B'\mat R^T, (\vec\omega_\body^{\core})^2} - \Io \nu\,\left(\vec\omega_\body^\core - \vec\omega_\body^\mantle\right)\,, \\
\dmat\pi_\mantle =&\; -g\comm{\vec k_\inertial, \vec\pi_\mantle} + \Ioc f_\core\comm{\mat R\mat B'\mat R^T, (\vec\omega_\body^\core)^2} + \Io \nu\,\left(\vec\omega_\body^\core - \vec\omega_\body^\mantle\right) \nonumber \\
&\;-3\Io n^2 \comm{\mat R\mat B'\mat R^T, \mat S}\,,
\label{eq.dotpim}
\end{align}
with now $\mat B' \equiv \mat R_3(\theta_p)(\mat B + \delta \mat B^\peri)\mat R_3(-\theta_p)$.
The second term in the right-hand side of $\dmat \pi_\core$ and $\dmat \pi_\mantle$ represents the conservative coupling between the core and the mantle generated by the pressure force of the liquid as it rotates while the third term accounts for the friction at the CMB. When adding the two equations of motion, we easily check that the total angular momentum $\vec \pi = \vec\pi_\core + \vec\pi_\mantle$ does satisfy the same equation of motion as in the single layer case (\ref{eq.dotpisolid}), as it should be.

To solve (\ref{eq.dotpic}) and (\ref{eq.dotpim}), we searched for a solution of the form $\mat R = \emi\vec i + \emj\vec j + \emk\vec k$ and $\vec\omega_\core = \omega_p\left(\ecj\vec i - \eci\vec j + \vec k\right)$. The calculation done with \TRIP{} shows that the in-plane obliquities $\emi$ and $\eci$ and the deviations from the Cassini plane $\emj$ and $\ecj$ again are decoupled from the mantle's rotation lag $\emk$ that is given by
\begin{equation}
\emk = (\kappa_{\zeta,s}+\kappa_{\zeta,c}+N_\mantle\emj)/(4\kappa_{22})\,.
\label{eq.emk}
\end{equation}
The other angles are solution the linearised system
\begin{equation}
\label{eq.syst2layers}
\mat A \begin{pmatrix}
\emi \\[0.3em]
\emj \\[0.3em]
\eci \\[0.3em]
\ecj
\end{pmatrix}
= - \begin{pmatrix}
N_\mantle \\[0.3em] 0 \\[0.3em] N_\core \\[0.3em] 0
\end{pmatrix} \,,
\end{equation}
where $\mat A = \mat A_\nu + \mat A_\omega + \mat A_\zeta$ is a $4\times4$ matrix given by
\begin{align}
\mat A_\nu =&\; \begin{bmatrix}
D_\mantle & F & E & -F \\[0.3em]
-F & D_\mantle & F & E \\[0.3em]
E & -F & D_\core & F \\[0.3em]
F & E & -F & D_\core
\end{bmatrix}\,, \\
\mat A_\omega =&\; -\kappa_\omega \begin{bmatrix}
\cos(2\varpi-2\Omega) &  \sin(2\varpi-2\Omega) & 0 & 0 \\[0.3em]
\sin(2\varpi-2\Omega) & -\cos(2\varpi-2\Omega) & 0 & 0 \\[0.3em]
0 & 0 & 0 & 0 \\[0.3em]
0 & 0 & 0 & 0 
\end{bmatrix}\,, \\
\mat A_\zeta =&\; \begin{bmatrix}
-K_1 + \kappa_{\zeta,c} & K_2 - \kappa_{\zeta,s} & K_3 & -K_4 \\[0.3em]
-K_2 + \kappa_{\zeta,s} &-K_1 + \kappa_{\zeta,c} & K_4 & K_3 \\[0.3em]
K_3 & -K_4 & -K_5 & K_6 \\[0.3em]
K_4 & K_3 & -K_6 & -K_5
\end{bmatrix}\,,
\end{align}
with the parameters
\begin{align}
E \equiv&\; MR^2 C_{20}^{\,\core}\,\omega_p\left(\omega_p + \dot\Omega\cos i\right)\,, \\
D_\mantle \equiv&\; \kappa + C_\mantle\omega_p\dot\Omega\cos i - MR^2C_{20}^{\,\core}\,\omega_p\dot\Omega\cos i - E\,,\\
D_\core \equiv&\;            C_\core\omega_p\dot\Omega\cos i   + MR^2C_{20}^{\,\core}\,\omega_p\dot\Omega\cos i - E\,, \\
F   \equiv&\; \Io \nu\,\omega_p\,, \\
N_\mantle \equiv&\; N - N_\core\,, \\
N_\core \equiv&\; \dot\Omega\sin i\left(C_\core  \omega_p+MR^2C_{20}^{\,\core}\omega_p\right)\,,
\end{align}
where $C_\core$ and $C_\mantle$ are the polar moment of inertia of the core and of the mantle, respectively,
and
\begin{align}
K_1 \equiv&\; \frac{1}{3}k_2q_rn^2MR^2\IfIc\left(3X^{-3,0}_0(e) + p^2\IfIc\right)\,,\\
K_2 \equiv&\; p\zeta K_1\,,\\
K_3 \equiv&\; \frac{1}{6}k_2q_rn^2MR^2\IfIc\left(3X^{-3,0}_0(e) + 2p^2\IfIc\right)\,,\\
K_4 \equiv&\; p\zeta K_3\,, \\
K_5 \equiv&\; \frac{1}{3}k_2q_r\omega_p^2MR^2\IfIc^2\,,\\
K_6 \equiv&\; p\zeta K_5\,,
\end{align}
where $\IfIc = \Ioc f_\core/\Io$. The coefficients $\kappa$, $\kappa_{22}$, and $\kappa_\omega$ have exactly the same expression as in the single layer model (see Eqs.~(\ref{eq.k20}-\ref{eq.kom})).
In this system of equations, we retained only the leading terms, disregarding those involving $\dot\Omega^2$ and smaller. A keen observer might notice that the system governed by Equations (\ref{eq.dotpic}) and (\ref{eq.dotpim}) contains six independent equations, whereas Equations (\ref{eq.emk}) and (\ref{eq.syst2layers}) together provide only five. Indeed, there is an additional equation that the system must satisfy, corresponding to the projection of $\dmat \pi_\core$ onto the unit vector $\vec k$. This equation has no solution under the assumption that the core's spin rate $\omega_\core$ equals the mantle's spin rate, $\omega_p$. However, if we allow the core spin rate to differ slightly from the mantle spin rate by a small amount $\delta\omega$, the final equation yields
\begin{equation}
\delta\omega \equiv \omega_\core - \omega_p = \frac{\left(C_\core + MR^2 C_{20}^{\,\core}\right)\ecj\omega_p\dot\Omega\sin i}{\Io \nu}\,.
\end{equation} 

In Equation (\ref{eq.syst2layers}), we decomposed the matrix $\mat A$ into three matrices, $\mat A_\nu$, $\mat A_\omega$, and $\mat A_\zeta$ to emphasise the distinct physical processes involved. The matrix $\mat A_\nu$ encapsulates the core-mantle interaction, namely the pressure torque (parameter $E$) and the viscous coupling (parameter $F$). The matrix $\mat A_\omega$ describes the apsidal precession of the orbit, as observed in \citet{Peale14} and detailed in \citet{Baland17}. The matrix $\mat A_\zeta$ accounts for the tidal deformation.

By neglecting apsidal precession and tidal deformation, and considering the frictionless regime ($F=0$) at the CMB, the system (\ref{eq.syst2layers}) decouples into two independent subsystems involving $(\emi, \eci)$ and $(\emj, \ecj)$, respectively. The $2\times2$ matrix with entries $D_\mantle$, $E$, $E$, and $D_\core$ is generically invertible, implying that $\emj$ and $\ecj$ must be zero. This signifies that, in this case, the Cassini state resides within the Cassini plane \citep{Boue20}. However, the inclusion of friction, apsidal precession, or tidal deformation couples $(\emi, \eci)$ and $(\emj, \ecj)$, leading to a deviation of the Cassini state from the Cassini plane.

In the following, solutions are obtained by inverting the system described in Equation (\ref{eq.syst2layers}). This analytical approximation holds valid as long as the quantities $\emi$, $\emj$, $\emk$, $\eci$, $\ecj$, and $\delta\omega/\omega_p$ are much smaller than one.

\section{Results}
\label{sec.results}
Here, we apply the model to Mercury where $p=3/2$, $\gamma_0 = 0$ and $\sigma_p = +1$. We briefly consider the solid case without molten core to compare our results with those of \citet{Baland17} and \citet{MacPherson22}. Then, we rapidly switch to the two-layer model for comparison with \citet{Peale14} and to explore a larger domain of the parameter space.

\subsection{Mercury as a solid body with no fluid core}
Using the parameters from \citet{Baland17}, we find the obliquity of Mercury in the Cassini plane to be $\varepsilon_\Omega = 2.026$ arcmin (Eq.~(\ref{eq.eps_Om_solid})) for the rigid case and $\varepsilon_\Omega^{k_2} = 2.031$ arcmin (Eq.~(\ref{eq.eps_Om_k2})) for the deformable case, in perfect agreement with \citet{Baland17}.

The radius of the circle described by the top of the spin axis, as the argument of the perihelion spans $180$ degrees \citep[see, e.g.][Fig.~8]{Baland17}, is given by our formula (\ref{eq.eps_om_solid}) as $\varepsilon_\omega = 0.898$ arcsec (hereafter, we refer to this motion as the apsidal circle). The value we obtain for $\varepsilon_\omega$ is slightly larger than the $0.868$ and $0.863$ arcsec found by \citet{Baland17} for the deformable and rigid cases, respectively.

However, we obtain a value closer to that of \citet{Baland17}, namely $\varepsilon_\omega = 0.865$ arcsec, when substituting $C_{20}$ and $C_{22}$ with $C_{20}+\delta C_{20}^\stat$ and $C_{22}+\delta C_{22}^\stat$, respectively. As explained earlier, we consider the $C_{20}$ and $C_{22}$ values measured with MESSENGER \citep{Mazarico14} to include the static component of their tidal deformation. Therefore, we favour the formula without the static corrections $\delta C_{20}^\stat$ and $\delta C_{22}^\stat$.

Lastly, for the deviation of the apsidal circle from the Cassini plane, we obtain $\varepsilon_\zeta = 0.0097$ arcsec, a value much smaller than its radius, in agreement with the green circle represented in \citep[Fig.~3]{Peale14}, whereas \citet{Baland17} found $0.995$ arcsec, a value approximately 100 times larger. As discussed previously, we believe the extreme value reported by \citet{Baland17} is due to incorrect projections of their equations of motion (59), which are not self consistent because of the lack of the rotation lag ($\varepsilon_3$ in our notation). This omission resulted in the unit vector $\hat s$ and its time derivative not being orthogonal.

Surprisingly, \citet{MacPherson22} reproduce the deviation angle $\varepsilon_\zeta$ from the Cassini plane, as found by \citet{Baland17}, using their Equation~(16), despite working in the mantle's equatorial plane, which by definition, is orthogonal to the figure axis. Therefore, the argument we presented against \citet{Baland17} does not apply in this case. The origin of the discrepancy between our result and that of \citet{MacPherson22} remains unclear.

\subsection{Influence of the molten core}
In \citet{Peale14}, the authors determined the equilibrium of Mercury's spin axis through numerical integration of equations of motion similar to our system (\ref{eq.eq_motion_2layers}), but also incorporating magnetic and topographic core-mantle couplings. One of our goals is to reproduce their results analytically, eliminating the need for numerical simulations. For easier comparison, we adopt their planetary parameters but retain the orbital parameters from \citet{Baland17}.

\begin{figure}
\begin{tikzpicture}
\path (0,0) node[below right=0, inner sep=0, outer sep=0] {%
\includegraphics[width=0.496\linewidth,trim=0 0 15 3,clip]{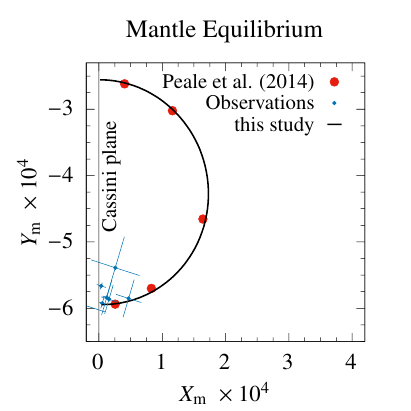}%
\includegraphics[width=0.496\linewidth,trim=0 0 15 3,clip]{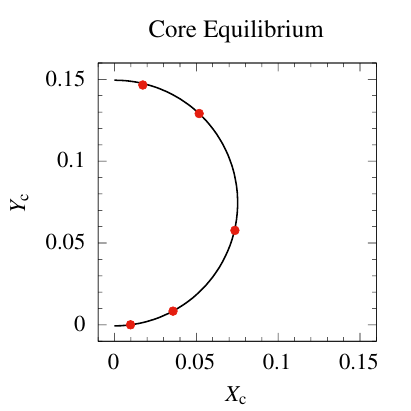}
};
\path (0,0) node[below right] {a)};
\path (0.5\linewidth,0) node[below right] {b)};
\end{tikzpicture}
\caption{\label{fig.peale}Projected equilibrium spin axes on the orbital plane for varying core-mantle viscous friction, assuming a spherical core. a) Mantle spin axis coordinates. b) Core spin axis coordinates. The vertical line labelled Cassini plane in panel a) is the intersection of the Cassini plane with the orbital plane. Red dots indicate coordinates from \citep[Table 2]{Peale14}. The observations are taken from \citep{Margot12,Mazarico14,Stark_rotation,Verma16,Genova19,Konopliv20,Bertone21}.}
\end{figure}

To validate our model, we reproduced Figure 3 from \citet{Peale14} in our Figure~\ref{fig.peale}, with red dots indicating the values from their Table 2. This figure shows the projection of the equilibrium spin axis onto the orbital plane for varying core-mantle viscous friction. Our model's agreement with their numerical solution is excellent. We also plot recent measurements of Mercury's spin pole orientation with their associated $1\sigma$ error bars. The model can explain the observed deviation from the Cassini plane, $\emj$, provided the tilt is towards positive abscissa ($X_\mantle > 0$). Among the seven measurements in Figure~\ref{fig.peale}, only that of \citet{Mazarico14} has a negative abscissa, with its error bar crossing the Cassini plane at the bottom of the figure. As noted in the introduction, the whole planet's moment of inertia, $C$, can be adjusted to fit the in-plane obliquity $\emi$ without affecting the longitude libration fit, provided the mantle's moment of inertia, $C_\mantle$, remains fixed \citep{Peale76}. Thus, fitting the spin axis orientation primarily involves determining the core's moment of inertia, $C_\core$, and the viscosity, parameterised by the relaxation time $\tau_\nu$.

Nevertheless, it is important to note that our model's leading-order computation of the planet's elastic deformation does not account for the fluid core spin obliquity. Including this effect could cause the viscous coupling to tilt the mantle spin axis towards $X_\mantle < 0$, as explained in \citep{MacPherson22}.

Figure 3 of \citet{Peale14} and our Figure~\ref{fig.peale} were obtained for a spherical core. However, \citet{Peale14} noted that the hydrostatic shape of the core would lead to a geometric flattening of $\varepsilon_f = 7.161\times10^{-5}$, corresponding to $f_\core=1.04$ in our notation. They also considered a non-hydrostatic case with twice the flattening, implying $f_\core=2.08$. With these values, \citet{Peale14} concluded that the spin axes of the core and mantle should be nearly aligned and close to the Cassini plane.

\begin{figure}
\begin{tikzpicture}
\path (0,0) node[below right=0, inner sep=0, outer sep=0] {%
\includegraphics[width=\linewidth,trim=0 0 0 38,clip]{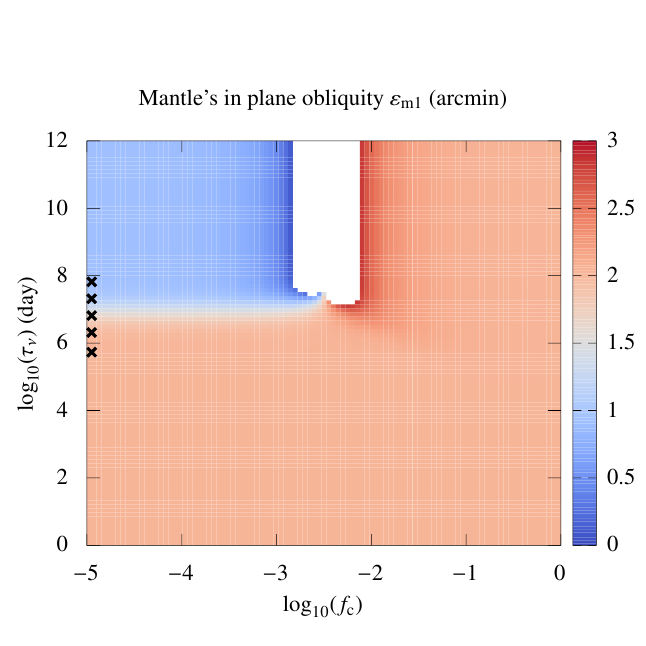}};
\path (0,-8cm) node[below right=0, inner sep=0, outer sep=0] {%
\includegraphics[width=\linewidth,trim=0 0 0 38,clip]{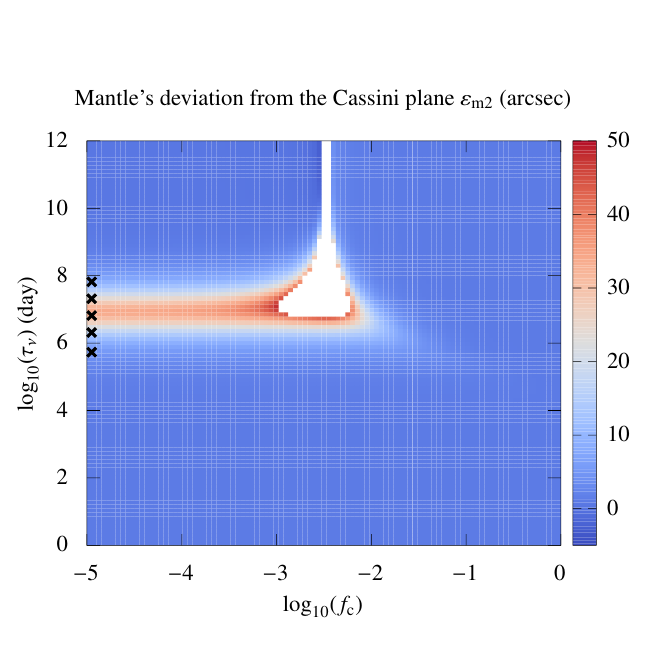}};
\path (0,0) node[below right] {a)};
\path (0,-8cm) node[below right] {b)};
\end{tikzpicture}
\caption{\label{fig.map}Mantle's in-plane obliquity $\emi$ (a) and deviation from the Cassini plane $\emj$ (b) as functions of the core flattening coefficient $f_\core$ and viscous relaxation time $\tau_\nu$. The white domains correspond to values that fall outside the colour scales. In these regions, the linear approximation is no longer valid and the equations of motions have to be solved treating the core's in-plane obliquity and angular velocity non-linearly. Black crosses indicate the parameters considered by \citet{Peale14}, shown as red dots in our Figure~\ref{fig.peale}.}
\end{figure}

To better understand this conclusion, we constructed a map of the mantle's in-plane obliquity, $\emi$, and deviation from the Cassini plane, $\emj$, as functions of the core flattening coefficient, $f_\core$, and viscous timescale $\tau_\nu$. The five cases considered by \citet{Peale14} in their Figure 3 are indicated by black crosses in our Fig.~\ref{fig.map}. The hydrostatic flattening of the core places Mercury slightly outside the map to the right. In Figure \ref{fig.map}a, we observe two plateaus: the light red area corresponds to $\emi \approx 2$ arcmin as observed, while the blue upper left corner shows a plateau at 1 arcmin. The ratio between these two values is exactly $C/C_\mantle$, implying that only the mantle is tilted in the blue region of the parameter space (Fig.~\ref{fig.map}a). Indeed, in the upper left corner, both flattening and viscous friction are low, decoupling the core from the mantle even on the nodal precession timescale. The transition occurring at $f_c\approx 0.005$ corresponds to a resonance of the precession angular speed with the nearly diurnal free wobble eigenfrequency \citep{Ragazzo22}. The parameters chosen by \citet{Peale14} ensure that the pressure torque is negligible ($f_\core \ll 1$), and the range of viscous timescales spans the transition between coupled and uncoupled core-mantle states. Figure~\ref{fig.map}b shows that this parameter space region exhibits significant deviation from the Cassini plane. As observed by \citet{Peale14}, adding the pressure torque ($f_\core$ close to one) causes the planet to behave like a solid body, reducing the deviation from the Cassini plane that is then solely induced by tidal deformation.

\begin{figure}
\includegraphics[width=\linewidth]{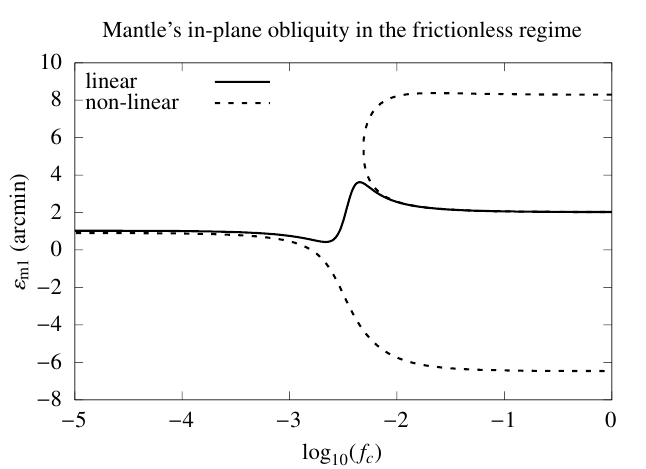}
\caption{\label{fig.nl}
Mantle's in-plane obliquity without viscous dissipation at the CMB ($\tau_\nu\to\infty$). The curve labelled 'linear' was computed using the present model and represents a transverse view of the top line in Fig.~\ref{fig.map}a. The non-linear curve is from \citep{Boue20}.}
\end{figure}

In Figures~\ref{fig.map}a and \ref{fig.map}b, the white domain corresponds to extreme solution values where the linear approximation used in this study fails. The parameter space studied by \citet{Boue20} corresponds to the upper line of the figure ($\tau_\nu\to\infty$ and a full range of $f_\core$). Although the linear model always provides a single solution, the non-linear solution obtained in the inviscid case exhibits a bifurcation (Fig.~\ref{fig.nl}). It may seem surprising that Mercury's obliquity shows non-linear effects despite its small value. This is due to the core reaching high obliquities of the order of 90 degrees, as shown in \citet{Peale14} and \citet{Boue20}. In this parameter space region, it is necessary at least to treat the core's in-plane obliquity, $\eci$, non-linearly. But as the core tilts, its angular speed significantly differs from the mantle's, requiring the addition of a new variable $\omega_\core$ to model this spin rate. And the relative angular speed, $\delta\omega = \omega_\core - \omega_p$, between the core and the mantle, must also be treated non-linearly.

\section{Conclusion}
\label{sec.conclusion}

In this paper, we present a new formulation of the classical model describing the rotation of Mercury, grounded in an advanced Lagrangian formalism developed by \citet{Ragazzo15} and \citet{Boue20}. This approach enables a comprehensive 3D solution for the mantle's orientation, including the phase in longitude at the perihelion, and in the future, will contribute to constraining its internal structure.

The final model takes into account a deformable mantle and a fluid ellipsoidal core, whose flattening imposes the conservative pressure torque. We also consider apsidal precession and dissipative effects associated with tidal interactions and friction between the core and mantle.

Firstly, we compare our results with those of \citet{Baland17} by treating Mercury initially as a rigid body and subsequently as a deformable body. We obtain similar values for the offset in the Cassini plane, with $2.026$ arcmin in the rigid case and $2.031$ arcmin in the deformable case. However, we find a radius of $0.898$ arcsec for the circle traced by the rotation axis during one perihelion precession, which is approximately 3\% larger than the value reported by \citet{Baland17}. This minor discrepancy might be attributed to a potential double-counting of the static component of the tidal deformation effects in their approach. More significantly, our calculation of the tidally induced deviation from the Cassini plane results in a tilt that is 100 times smaller than the value reported by \citet{Baland17}. We suggest that the out-of-Cassini-plane offset may have been overestimated in their study due to an incorrect projection of the angular momentum equations onto the invariant plane. 

Secondly, we obtain for the complete model the derivation of analytical equilibrium equations for Mercury’s rotation axis. We reproduce the equilibrium positions obtained numerically by \citet{Peale14}, which are compatible with the latest observations of Mercury's rotation axis. We focus particularly on the influence of core flattening and CMB friction on both in-plane and out-of-plane offsets from the Cassini plane. We confirm the result of \citet{Margot12}, which concluded that if the core and mantle are too weakly coupled, it becomes impossible to explain the observed in-plane offset. As in \citet{Peale14}, we show that for a nearly spherical core, friction plays a dominant role, even more so than in the case of a more flattened core. This leads to a regime where the out-of-Cassini-plane offset becomes excessively large, rendering it incompatible with observations. 

Finally, we identify a parameter space region where the linear approximation in Mercury's orientation breaks down. This region is characterised by weak core-mantle friction and intermediate core flattening. In this regime, a bifurcation occurs, as described in the frictionless limit by \citet{Boue20}, when the core angular momentum tilts beyond 30 degrees. Additionally, the core's angular speed, forced by core-mantle friction, should significantly decrease, in contradiction with our assumption that the core and mantle angular speeds are equal.

Nevertheless, comparing our results with those of \citet{Peale14} – which are not subject to the linear approximation – shows that a significant out-of-Cassini-plane deviation can still occur within the validity range of our model. This scenario assumes a nearly spherical core and a viscous relaxation time, driven by core-mantle friction, of the order of $10^7$ days. In this situation, tidal deformation plays a minimal role compared to the core flattening and CMB friction.

Future work is still needed to constrain Mercury's interior structure using the upcoming data from the BepiColombo space mission. This work will focus on incorporating a solid inner core, a realistic fluid rotation model for the outer core, a more complex rheological model, and a more refined orbital motion that goes beyond the uniform precession at constant inclination and eccentricity.

\begin{acknowledgements}
We would like to thank Benoît Noyelles, whose valuable comments helped to improve the quality of the paper.
C.~R. is partially supported by FAPESP grant 2023/07076-4.
\end{acknowledgements}

\bibliographystyle{aa}
\bibliography{mercury}
\end{document}